\documentclass[journal]{IEEEtran}

\ifCLASSINFOpdf
\else
\fi

\usepackage{graphicx}
\usepackage{color}
\usepackage{placeins}
\usepackage{float}
\usepackage{tabularx,colortbl}
\usepackage{gensymb}
\usepackage{cite}
\usepackage{amsmath,amssymb}
\usepackage{listings}
\usepackage{caption}
\usepackage{lipsum}
\usepackage{mwe}
\usepackage{breqn}
\usepackage{hhline}
\begin{document}

\title{High-Accuracy Multi-Node Ranging For Coherent Distributed Antenna Arrays}
%\title{Two-Step Numerical Optimization of Linear Arrays With Extreme Sparsity Using a Genetic Algorithm and Particle Swarm Optimization}

%\author{\IEEEauthorblockN{Sean M. Ellison}
%\IEEEauthorblockA{ECE, Michigan State University\\
%Email: elliso65@msu.edu}
%\author{Sean M. Ellison,~\IEEEmembership{Student Member,~IEEE} and Jeffrey A. Nanzer,~\IEEEmembership{Senior Member,~IEEE}% <-this % stops a space

\author{Sean M. Ellison,~\IEEEmembership{Student Member,~IEEE} and Jeffrey A. Nanzer,~\IEEEmembership{Senior Member,~IEEE}
\thanks{This material is based upon work supported by the Air Force Research Laboratory (contract number FA8750-17-C-0182) and by The Defense Advanced Research Projects Agency (grant number N66001-17-1-4045). The views, opinions, and/or findings contained in this article are those of the author and should not be interpreted as representing the official views or policies, either expressed or implied, of the Defense Advanced Research Projects Agency or the Department of Defense.}
\thanks{The authors are with the Department of Electrical and Computer Engineering, Michigan State University, East Lansing, MI 48824 USA (email: elliso65@msu.edu, nanzer@msu.edu).}
}% <-this % stops a space

% <-this % stops a space

%\markboth{IEEE}%
%{Shell \MakeLowercase{\textit{et al.}}: Bare Demo of IEEEtran.cls for Journals}

\maketitle

\bstctlcite{IEEEexample:BSTcontrol} % adds et. al. to bibliography

\begin{abstract}
The design and experimental implementation of a waveform for high-accuracy inter-node ranging in a coherent distributed antenna array is presented. Based on a spectrally-sparse high-accuracy ranging waveform, the presented multi-frequency waveform enables high-accuracy ranging between multiple nodes in an array simultaneously, without interference. The waveform is based on a unique time-frequency duplexing approach combining a stepped-frequency waveform with different step cycles per node pair. The waveform also inherently includes beneficial disambiguation properties. The ambiguity function of the waveform is derived, and theoretical bounds on the ranging accuracy are obtained. Measurements were conducted in software-defined radio-based nodes in a three-element distributed array, demonstrating high-accuracy unambiguous ranging between two slave nodes and one master node.
\end{abstract}

%\IEEEoverridecommandlockouts


\IEEEpeerreviewmaketitle

%%%%%%%%%%%%%%%%%%%%%%%%%%%%%%%%%%%%%%%%%%%%%%%%%%%%%%%%%%%%%%%%%%%%%%%%%%%%%%%%%%%%%%%%%%%%%%%
\section{Introduction}
Advancements in the capabilities of wireless systems strongly depend on the achievable gain, power, and resolution of the system, and the ability to scale these metrics. With traditional single-platform systems, improvements in such aspects for radar, remote sensing, and communications requires modifications of the devices in the system, the system efficiency, or the antenna aperture size, all of which can represent significant cost drivers. To overcome the challenge of continually upgrading single-system capabilities, there is growing interest in distributed antenna array systems that can be easily scaled by adding small, inexpensive nodes to the array, thereby increasing the capabilities of the overall array system through array gain and/or larger array aperture area \cite{7803582,788373,6206634}.
%,5136190}. 
Applications of such disaggregated arrays include cubesat swarm operations for remote sensing, drone constellations for soil moisture measurements, and distributed communications for greater throughput and higher-reliability connections, among others.

Coherent distributed arrays are a specific subset of distributed arrays in which platforms are coordinated on the level of radio frequency phase to enable phased-array beamforming \cite{mudumbai2009distributed,8058723}. To support distributed coherent operations, the wireless systems on each platform must be phase-aligned to ensure that signals add constructively\cite{chiu2010dynamic, musch2005phase}, and time aligned so that symbol information is sufficient overlapped at the target destination \cite{chatterjee2018effects, shen2011time,6731859,7147237}. Of these two, the phase alignment is the more challenging task due to the high-precision required to align the carrier phases. Previous works have performed this task using a feedback methodology where array position information is sent to a the coherence location where corrections in position or phase can be calculated \cite{876517,7771912,4202181,4785387}. This generally restricts array space to where reliable positioning information is available. The work presented here focuses on an open-loop array where no feedback from the destination is assumed \cite{995421,4305509,512816}. Although open-loop arrays are a considerably more daunting task to develop, there are huge versatility benefits due to the ability to arbitrarily steer beams. For this process to take place not only must the transmitting oscillators be phased locked, but the relative positions of the platforms must be measured to within a fraction of a wavelength. 

This work focuses on the task of achieving high-accuracy inter-node ranging. Previously, it was shown that in order to achieve a high probability of a high level of coherent gain in a distributed array, ranging accuracies on the order of a fraction of a wavelength of the transmitted waveform are required \cite{7803582}. Specifically, to achieve no more than a 0.5 dB reduction in coherent gain with a probability of 90\%, the relative ranges must be measured to within an accuracy of $\lambda/15$, which for microwave systems is on the order of centimeters. Furthermore, in dynamic distributed arrays, this accuracy must be achieved quickly, before the nodes move out of the coherence time of the channel. While optical systems can achieve this level of accuracy, such systems are not scalable, requiring accurate pointing and tracking for each node connection \cite{4026149,189658}. Prior work demonstrated that a spectrally-sparse two-tone microwave signal achieves optimal ranging accuracy and is feasible to broadcast in the microwave band \cite{7118937,7304876}, however this waveform is highly ambiguous, and has been demonstrated only between two nodes. For larger arrays, it is necessary to achieve high-accuracy ranging between multiple nodes simultaneously.

In this paper, a scalable, high-accuracy ranging waveform operating at microwave frequencies is developed and demonstrated. Based on the high-accuracy two-tone waveform, the presented signal format is modified to enable duplexing in the time-frequency domain. Based on a stepped-frequency waveform, the stepping cycle can be modified to match specific node pairs, and enables simultaneous ranging operation without interference. Furthermore, the waveform has beneficial disambiguation properties: as the number of steps is increased, the ambiguities immediately adjacent to the desired range response are nulled in a linear fashion. Section II describes the overall scalability approach, following which Section III reviews the two-tone waveform, a practical implementation of the two-tone waveform in a pulsed format, and the new stepped-frequency waveform. The theoretical accuracy of the waveform is derived, along with the ambiguity function. Measurements of the accuracy demonstrate that the stepped-frequency waveform is not appreciably different than the pulsed two-tone waveform. Section IV describes a multi-node measurement system implemented using software-defined radios, where the stepped-frequency waveform is tested in a three-node unambiguous ranging format. Measurements of inter-node absolute range and accuracy demonstrate the feasibility of the design.

%%%%%%%%%%%%%%%%%%%%%%%%%%%%%%%%%%%%%%%%%%%%%%%%%%%%%%%%%%%%%%%%%%%%%%%%%%%%%%%%%%%
\section{Frequency Domain Multiplexing}
Coherent distributed arrays consisting of large numbers of nodes generally necessitate some multiplexing approach to enable inter-node ranging between multiple node pairs. While this may be accomplished using time-domain multiplexing, where the range between each node is done in sequence, with mobile nodes there will inherently be some time limitations after which the motion of the nodes, either from intentional movement or inherent platform vibration, causes the measurement to no longer be sufficiently accurate. It is thus preferable to begin with a frequency-domain multiplexing (FDM) approach where multiple measurements can be accomplished simultaneously, after which additional time-multiplexing may be included. 

Frequency-domain multiplexing contains some inherent time-domain information, since the bandwidth of the channels used for each node pair will dictate limits on the waveform length. If a half-duplex system is available then there is a limitation on the minimum detectable time delay that can be measured due to bandwidth constraints such that $T_{min}=2/BW$ where $BW$ is the total available bandwidth. With this minimum available time the minimum detectable range can be calculated as $R_{min}=c/T_{min}$ where $c$ is the speed of light. This $T_{min}$ will now be the lower limit of the pulse length for a half duplex system. For a full duplex system there is no minimum range limitation due to the ability to transmit and receive simultaneously. The dwell time for each transmission measurement will be at least $2T$ where $T$ is the pulse length which is fixed to the max dimension of the array such that $T=R_{max}/c$ so there is sufficient time for the signal to travel the full extent of the array. Therefore the minimum waveform length is defined by the type of system that is available and by the dimensions of the array. It is to be noted that these are minimum temporal limits and in practice waveforms will be much longer than this mostly due to hardware limitations.

Once the limit on temporal length of the waveform is determined, the bandwidth that each pulse occupies can be estimated as $\Delta f=1/T$. Using FDM  the number of simultaneous connections can be estimated as $m=BW/\Delta f$ \cite{1096357}. We can now assume that $m$ connections can be made in $2T$ accounting for two way propagation. Given metrics relative to the coherence time of the channel between the elements, such as the vibration profile of the platforms, a desired update rate $\Delta t$ can be chosen. The total number of connectable nodes can then be derived from the number of $m$ connections that can be made in $\Delta t$. An estimation of this number can be expressed as
\begin{equation}
	N=\frac{BW\Delta t}{2}
\end{equation}

\begin{figure} [t!]
	\begin{center}
		\includegraphics[width=3.5in]{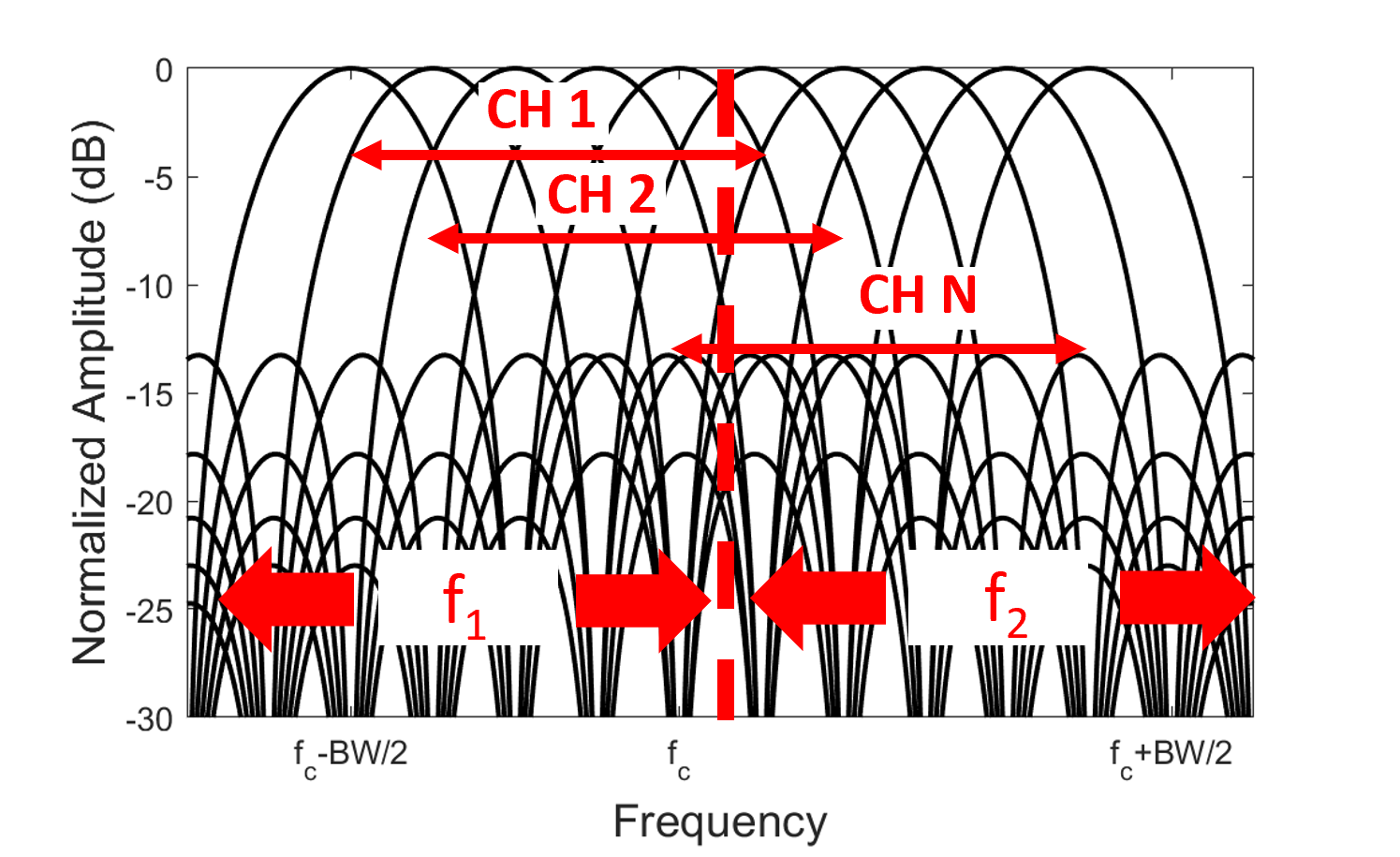}
	\end{center}
	\caption{Channel splitting using an FDM method}
	\label{fig:channel}
\end{figure}
Once the domain is divided into bands, two bands are designated for a given node pair to form a channel. Channels are given two bands due to support the spectral sparsity of the waveform, as described in the next section. To prevent measurement bias, the channels are sequentially organized such that the tone separation of each channel remains constant (see Fig. \ref{fig:channel}).

%%%%%%%%%%%%%%%%%%%%%%%%%%%%%%%%%%%%%%%%%%%%%%%%%%%%%%%%%%%%%%%%%%%%%%%%%%%%%%%%%%%%%%%%%%%%%%%
\section{Waveform Design}
The multi-node high-accuracy ranging waveform is based on a spectrally-sparse, two-tone waveform. In this section, the ideal two-tone waveform is first derived and discussed, followed by a practical implementation of the two-tone waveform in a pulsed form. Following this, the multi-node stepped-frequency waveform is derived and discussed.

\subsection{Two-Tone Waveform}
To have adequate phase alignment of the elements to support distributed beamforming, the relative positions of the elements need to be found within a fraction of a wavelength. The measurement accuracy that can be achieved is highly dependent on the type of waveform that is being used. A typical radar ranging waveform can be modeled as
\begin{equation}
	s_r(t)=\alpha g(t-\tau)e^{j2\pi f_d t}+w(t)
\end{equation}
where $\alpha$ is the amplitude, $\tau$ is the time delay, $f_d$ is the Doppler shift, and $w(t)$ is the noise, which for this work is considered to be additive Gaussian white noise. Using this form, the variance of the error in the measurement is given by the Cramer-Rao Lower Bound (CRLB) \cite{van2004detection}
\begin{dmath} \label{eq:2}
	\text{var}(\hat{\tau}-\tau)\geq\frac{N_0}{2|\alpha|^2}\Big(\int\Big|\frac{\partial }{\partial \tau}g(t-\tau)\Big|^2 dt-\frac{1}{E_s}\Big|\int \bigg(\frac{\partial }{\partial\tau}g(t-\tau)\bigg)^*g(t-\tau)dt\Big|^2\Big)^{-1}
\end{dmath}
where $N_0$ is the noise spectral density, $E_s$ is the signal energy, and $\frac{N_0}{|\alpha|^2}= \frac{1}{SNR}$ where SNR is the signal-to-noise ratio of the measurement. The two integrals represent the first and second moment of the frequency spectrum respectively,
\begin{align}
	\Big|\int \bigg(\frac{\partial }{\partial\tau}g(t-\tau)\bigg)^*g(t-\tau)dt\Big|^2&=\mu_f^2\\
	\int\Big|\frac{\partial }{\partial \tau}g(t-\tau)\Big|^2 dt&=\zeta_f^2
\end{align}
where $\mu_f$ is mean frequency and $\zeta_f$ is the mean bandwidth. The measurement variance is then
\begin{dmath} \label{eq:5}
	\text{var}(\hat{\tau}-\tau)\geq\frac{N_0}{2|\alpha|^2\big(\zeta_f^2-\frac{1}{E_s}\mu_f^2\big)}
\end{dmath}
The waveforms considered here, as is the case in general with most radar waveforms, are symmetric in the two sided frequency spectrum (i.e. $\mu^2_f=0$). Therefore, the variance estimate is inversely proportional to the second moment of the frequency spectrum.

Because the measurement variance is inversely proportional to the mean-square bandwidth, the variance can be minimized by maximizing this quantity. This is done by concentrating the waveform energy in the edges of the band (i.e. two-tone waveform). Evaluating the mean-square bandwidth of a two-tone signal with frequencies separated by $\Delta f$ yields a measurement variance given by
\begin{dmath} \label{eq:6}
	\text{var}(\hat{\tau}-\tau)\geq\frac{N_0}{2\pi^2|\alpha|^2\Delta f^2}
\end{dmath}
This represents the lowest variance that can be achieved; all other waveform types will serve to increase the estimation variance, resulting in a reduction in performance.

For maximum performance a continuous wave is needed due to the representation in the Fourier domain which will result a delta function at the designated frequencies, yielding a perfect two-tone signal. In practice, the signal will necessarily be time-limited. To assess the effects of this time limitation, a two-tone waveform modulated by a square temporal pulse is analyzed here. In the frequency domain, this will create sinc functions at the designated two frequencies whose bandwidth is inversely proportional to the pulse duration. This will result in a decrease of the mean squared bandwidth and in turn will raise the minimum obtainable variance. As long as the pulse bandwidth is small compared to the frequency separation of the two tones, this reduction in accuracy will have little effect on the variance.

The pulsed two-tone waveform can be scaled by using frequency multiplexing by choosing $f_1$ and $f_2$ to be the two carrier frequencies of the two-tone signal and adding additional channels by increasing both $f_1$ and $f_2$ by a frequency increment $\delta f$ until the available bandwidth is filled. In this setup each channel consisting of two frequency bands (see Fig. \ref{fig:channel}) will service one connection between a pair of nodes.

While the variance of the waveform gives a measure of stability on a measurement by measurement basis, it is also important to observe the measurement resolution (both in range and Doppler) to indicate how ambiguous the measurement results might be. Ambiguous results can lead to incorrect range estimates. The ambiguity function is often used to evaluate the performance of a waveform in both delay and Doppler. The time domain representation of a pulsed two-tone waveform (PTTW) which can be expressed as
\begin{equation} 
	S(t)=\text{rect}\Big(\frac{t}{T}\Big)\Big(e^{j2\pi f_1 t}+e^{j2\pi f_2 t}\Big)
\end{equation}
where rect($\cdot$) is the rectangular pulse function and $T$ is the pulse duration. The ambiguity function is the auto-correlation of the transmitted signal multiplied by a Doppler shifted version, given by
\begin{equation} \label{eq:8}
	AF(t,f_D)=\int_{-\infty}^{\infty}S^*(\tau-t)S(\tau)e^{j2\pi f_D \tau}d\tau
\end{equation}
The ambiguity function can be derived for the PTTW as
\begin{dmath}
\bigg|AF(t,f_D)\bigg|=\Bigg|\big(T-|t|\big)\Bigg(\bigg(e^{j2\pi f_1t}+e^{j2\pi f_2t}\bigg)\\
\times\text{sinc}\bigg(\pi f_D(T-|t|)\bigg)\\
+e^{j\pi (f_2+f_1)t}\text{sinc}\bigg(\pi (f_D\pm f_2\mp f_1)(T-|t|)\bigg)\Bigg)\Bigg|
\end{dmath}

\begin{figure} [t!]
	\centering
	\begin{minipage}[t]{0.24\textwidth}
		\centering
		\includegraphics[width=\textwidth]{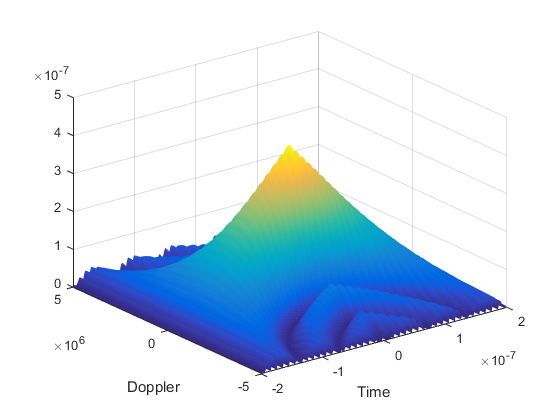}
	\end{minipage}
	\hfill
	\begin{minipage}[t]{0.24\textwidth}
		\centering
		\includegraphics[width=\textwidth]{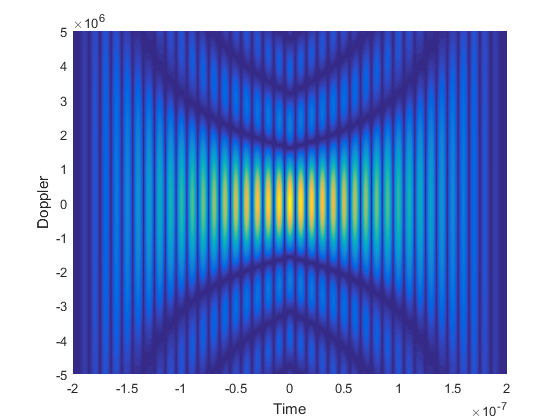}
	\end{minipage}
		\caption{Pulsed two-tone waveform ambiguity function}
		\label{fig:ptt_af}
\end{figure}
An image of the ambiguity function for a PTTW can be seen in Fig. \ref{fig:ptt_af}. It can be seen there is a repeated lobing pattern in the time domain. This arises from the beat frequency between $f_1$ and $f_2$. In the frequency domain the response is a sinc function which stems from the overlaying rectangular pulse modulation.

While having the benefit of increased measurement accuracy, a PTTW waveform has a drawback which is that the measurement is highly ambiguous due to the large number of lobes in the temporal response. It is typically challenging to track the correct peak lobe especially in the presence of noise. Previous works have used an interweaved wideband pulse with bandwidth equal to $\Delta f/4$ to have effective peak tracking. While this brute force method is effective, this paper will explore a more elegant method of disambiguation through waveform design alone.

\subsection{Stepped-Frequency Waveform}
\begin{figure} [t!]
	\centering
	\begin{minipage}[t]{0.24\textwidth}
		\centering
		\includegraphics[width=\textwidth]{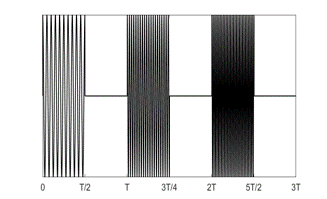}
	\end{minipage}
	\hfill
	\begin{minipage}[t]{0.24\textwidth}
		\centering
		\includegraphics[width=\textwidth]{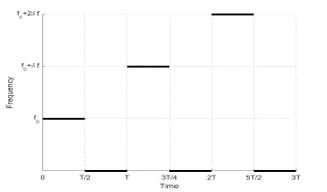}
	\end{minipage}
		\caption{Stepped frequency waveform time spectrum (left) and time frequency spectrum (right)}
		\label{fig:sfw}
\end{figure}

The stepped-frequency waveform (SFW) was designed as a pulse compression method that achieves effective wideband measurements through several consecutive pulses with narrow instantaneous bandwidths \cite{4148599}. This is done by having a carrier frequency $f_c$ that is monotonically increased by an increment $\delta f$ for every successive pulse. This waveform is also capable of acheiving high range resolution. The waveform can be modeled as 
\begin{equation}
	S(t)=\frac{1}{\sqrt{N}}\sum_{n=0}^{N-1}\text{rect}\bigg(\frac{t-nT_r}{T}\bigg)e^{j2\pi f_c t}e^{j2\pi n\delta f t}
\end{equation}
An image of the waveform model, with a 50\%  temporal duty cycle, in both the time domain and the time-frequency domain can be seen in Fig. \ref{fig:sfw}.

\begin{figure} [t!]
	\centering
	\begin{minipage}[t]{0.24\textwidth}
		\centering
		\includegraphics[width=\textwidth]{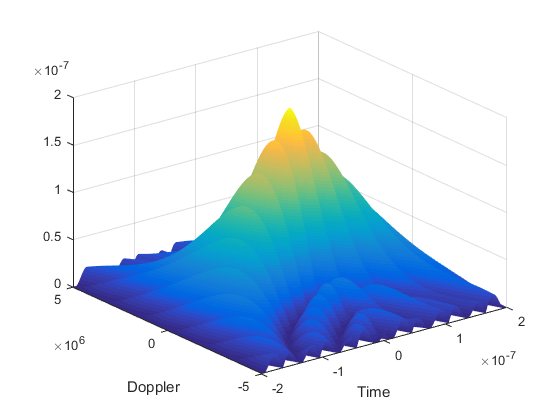}
	\end{minipage}
	\hfill
	\begin{minipage}[t]{0.24\textwidth}
		\centering
		\includegraphics[width=\textwidth]{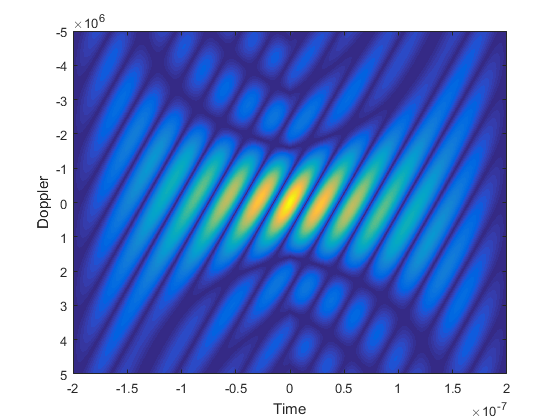}
	\end{minipage}
		\caption{Stepped-frequency waveform ambiguity function}
		\label{fig:sfw_amb}
\end{figure}

There are two main advantages to using a SFW: the range resolution is increased while maintaining narrow instantaneous bandwidth \cite{4800321}, and the next highest sidelobe of the matched filter is shifted by $t=\frac{1}{\delta f}$. This can be observed from the ambiguity function of the SFW \cite{574466}, given by
\begin{dmath}
	\left|AF(t,f_D)\right|=\left|\left(T-|t|\right)\text{sinc}\left(\pi f_D(T-|t|)\right)\right|\\
	\times\left|\frac{\sin\left(\pi N(\delta f t+f_DT_r)\right)}{N\sin\left(\pi(\delta ft+f_DT_r)\right)}\right|\\
\end{dmath}
When looking at the $f_D=0$ cut, this function will be maximized at every $t=\frac{n}{\delta f}$ and therefore the next highest sidelobe will be at $t=\frac{1}{\delta f}$. Considering the nulls of the zero Doppler cut, which occurs at $t=\pm\frac{1}{N\delta f}$, it can now easily be seen that the lobe beamwidth is inversely proportional to the number of pulses $N$. A plot of the SFW ambiguity function can be see in Fig. \ref{fig:sfw_amb}.

The disambiguation properties of this waveform can clearly be seen to be inversely proportional to the frequency step $\delta f$. Therefore large frequency steps will result in unambiguous measurements but this will require a large amount of bandwidth. Although this waveform has beneficial disambiguation properties, there is one main drawback, which is that to obtain the resolution of that of the PTTW in Eq. \ref{eq:5}, it would take a large number of pulses $N$ and a long transmit/receive time to capture all the pulses.

\subsection{Two-Tone Stepped-Frequency Waveform}
To combine the advantage of both the resolution of the PTTW and the disambiguation ability of the SFW, a two-tone stepped-frequency waveform (TTSFW) was developed. This was done by first choosing the individual pulse bandwidth, $f_1$ and $f_2$, then monotonically increasing these frequencies by $\delta f$ keeping the bandwidth of every pulse the same. 

\begin{figure} [t!]
	\centering
	\begin{minipage}[t]{0.24\textwidth}
		\centering
		\includegraphics[width=\textwidth]{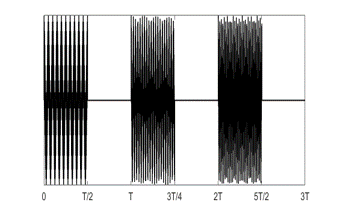}
	\end{minipage}
	\hfill
	\begin{minipage}[t]{0.24\textwidth}
		\centering
		\includegraphics[width=\textwidth]{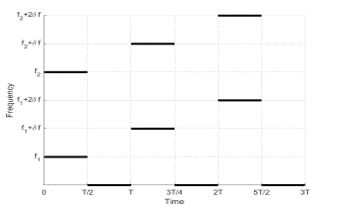}
	\end{minipage}
		\caption{Two-tone stepped-frequency waveform time spectrum (left) and time frequency spectrum (right)}
		\label{fig:ttsfw}
\end{figure}

This waveform can be scaled similarly to that of the PTTW but instead of having only one pulse there the waveform includes an arbitrary N pulses, essentially stitching N different PTTW pulses into one waveform. Since more resources are being used per node connection it is intuitive that the scalability would decrease, however this is not the case. In fact the same waveform with N pulses can be used to service N connections. This is done by shifting the order of pulses (frequency steps) so that there are N unique waveforms utilizing the same pulses, each in a different order. The TTSFW signal can be modeled as
\begin{equation}
	S(t)=\frac{1}{\sqrt{N}}\sum_{n=0}^{N-1}\text{rect}\bigg(\frac{t-nT_r}{T}\bigg)\bigg(e^{j2\pi f_1t}+e^{j2\pi f_2t}\bigg)e^{j2\pi n\delta ft}
\end{equation}
An image of the time frequency spectrum along with the waveform in the time domain can be seen in Fig. \ref{fig:ttsfw}. 

The CRLB can be derived for this waveform using Eq. \ref{eq:5}, which yields
\begin{equation} \label{eq:12}
	\text{var}(\hat{\tau}-\tau)\geq\frac{N_0}{2|\alpha|^2\big(\pi^2\Delta f^2+\frac{(2\pi\delta f)^2}{N}\sum_{n=0}^{N-1}n^2\big)}
\end{equation}
Using the scalability approach described above, the frequency step and pulse bandwidth can be expressed in terms of the full system bandwidth as $\delta f=\frac{BW}{2N-1}$ and $\Delta f=N\delta f=\frac{NBW}{2N-1}$. The mean squared bandwidth from Eq. \ref{eq:12} can be represented as
\begin{equation}\label{eq15}
	\zeta_f^2=\pi^2\bigg(\frac{BW}{2-\frac{1}{N}}\bigg)^2+\frac{(2\pi BW)^2}{N(4N^2+4N+1)}\sum_{n=0}^{N-1}n^2
\end{equation}

\begin{figure} [t!]
	\begin{center}
		\includegraphics[width=3.5in]{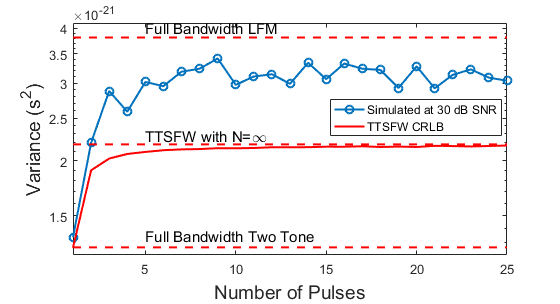}
	\end{center}
	\caption{The bounds and simulation of TTSFW vs. number of pulse with 4 MHz of bandwidth total bandwidth and 30 dB preprocessing SNR}
	\label{fig:bounds}
\end{figure}

\begin{figure} [t!]
	\begin{center}
		\includegraphics[width=3.5in]{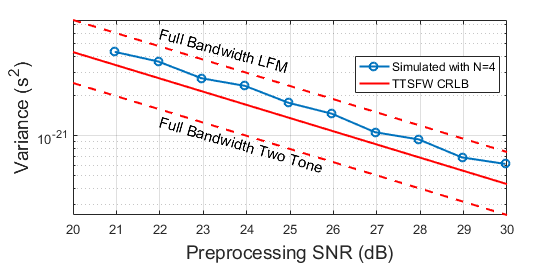}
	\end{center}
	\caption{The bounds and simulation of TTSFW vs. SNR with the pulse number fixed to $N=4$.}
	\label{fig:bounds2}
\end{figure}

Looking at the limits of this function, if $N=1$ then the second term will become zero and the first term will become $\pi^2 BW^2$. This is exactly the result found in Eq. \ref{eq:6} in the two-tone case. Looking at the opposite bound, when $N=\infty$, the the first term will approach
\begin{equation}\label{eq16}
\lim_{N\rightarrow\infty}\Bigg(\pi^2\bigg(\frac{BW}{2-\frac{1}{N}}\bigg)^2\Bigg)=\frac{(\pi BW)^2}{4}
\end{equation}
and the second term
\begin{equation}\label{eq17}
(2\pi BW)^2\lim_{N\rightarrow\infty}\bigg(\frac{1}{N(4N^2+4N+1)}\sum_{n=0}^{N-1}n^2\bigg)=\frac{(\pi BW)^2}{3}
\end{equation}
The term in Eq. \ref{eq16} is the two tone case when the individual pulse bandwidth is $BW/2$. This follows the scalability approach described in Section II when there is a large number $N$ (see Fig. \ref{fig:channel}). The term in Eq. \ref{eq17} is exactly the bound that is obtained by using a linear frequency modulated (LFM) signal with equal amplitude across the frequency band $BW$ \cite{Skolnik1962}. It can be deduced that the first term in Eq. \ref{eq15} is a measure of the mean squared bandwidth of the individual pulses and the second term is a measure of the mean squared bandwidth of the overall waveform. A plot of the TTSFW's accuracy vs. number of pulses used can be seen in Fig. \ref{fig:bounds}. From this figure, the best obtainable bound is for when $N=1$ which is the two-tone case. It then increases in a logarithmic fashion until the upper bound  of $(\pi BW)^2/4+(\pi BW)^2/3$ is reached at approximately $N=20$. To test the robustness of the waveform as the number of pulses increases, 1000 Monte Carlo simulations were preformed with 4 MHz of total bandwidth at approximately 30 dB preprocessing SNR and duration of 500 $\mu$s sampled at 25 MHz. This is to emulate what is achievable by the measurements in the following section. The simulated results as a function of the number of pulses, $N$, can also be seen on the Fig. \ref{fig:bounds} while a plot of the simulation with a fixed $N$ and variable SNR can be seen in Fig. \ref{fig:bounds2}.

The disambiguation properties of the waveform can be observed from the ambiguity function of the TTSFW
\begin{dmath}
	\Big|AF(t,f_D)\Big|=\Bigg|\big(T-|t|\big)\Bigg[\Bigg(e^{j2\pi f_1t}+e^{j2\pi f_2t}\Bigg)\\
	\times\text{sinc}\bigg(\pi f_D(T-|t|)\bigg)\frac{\sin\bigg(N\pi\big(\delta ft+f_DT_r\big)\bigg)}{N\sin\bigg(\pi\big(\delta ft+f_DT_r\big)\bigg)}\\
	+e^{j\pi (f_2+f_1)t}\text{sinc}\bigg( \pi (f_D\pm f_2\mp f_1)(T-|t|)\bigg)\\
	\times\frac{\sin\bigg(N\pi\big(\delta ft+(f_D\pm f_2\mp f_1)T_r\big)\bigg)}{N\sin\bigg(\pi\big(\delta ft+(f_D\pm f_2\mp f_1)T_r\big)\bigg)}\Bigg]\Bigg|\\
\end{dmath}
\begin{figure} [t!]
	\centering
	\begin{minipage}[t]{0.24\textwidth}
		\centering
		\includegraphics[width=\textwidth]{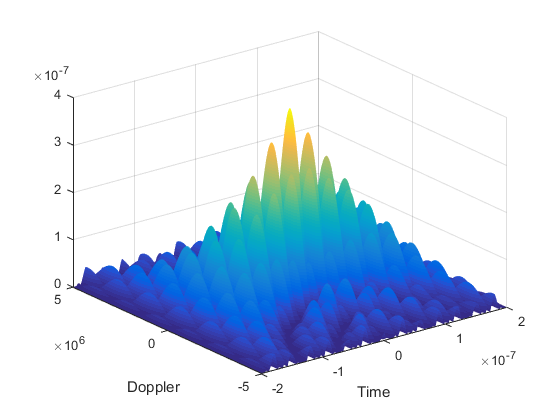}
	\end{minipage}
	\hfill
	\begin{minipage}[t]{0.24\textwidth}
		\centering
		\includegraphics[width=\textwidth]{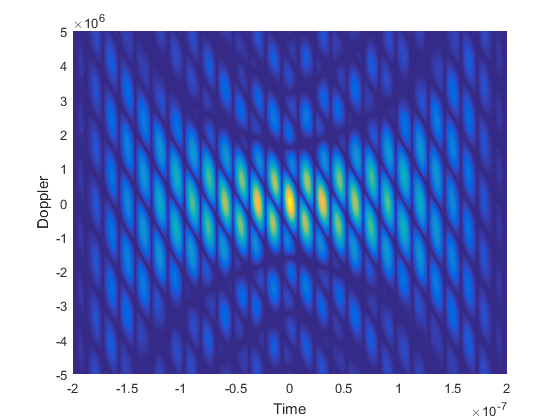}
	\end{minipage}
		\caption{Two-tone stepped-frequency waveform ambiguity function}
		\label{fig:ttsfw_amb}
\end{figure}
A plot of this function can be seen in Fig. \ref{fig:ttsfw_amb}. Looking at the $f_D=0$ cut of the ambiguity function, this function is maximized at every $t=\frac{nN}{\Delta f}$. Since the lobing pattern of the PTTW occurs at every $t=\frac{n}{\Delta f}$ the disambiguation properties of this waveform are now apparent: for an $N$ pulse system, there will be $N-1$ consecutive lobes notched out of the matched filter output of the PTTW with the same $\Delta f$. It is important to note, however, that the spectrally sparse nature of the TTSFW waveform, along with the ability to simultaneously transmit multiple waveforms simultaneously, makes the TTSFW more beneficial to implement than a traditional LFM. A comparison of the beneficial attributes of the derived waveforms can be seen in Table \ref{tab:table1}.
\begin{table}[t!]
\caption{Comparison of discussed waveform attributes}
%\begin{tabular}{|C|C|C|C|}
\begin{tabular}{c|ccc}
\hline
\textbf{
    \begin{tabular}{@{}c@{}}Rank based\\ on CRLB \\ (lowest to highest) \end{tabular}
	}	
   &\textbf{Waveform} & \textbf{Scalable}  & \textbf{Unambiguous}\\ \hline\hline
 1 &\begin{tabular}{@{}c@{}}Full Bandwidth \\ Two Tone \end{tabular}
	 & No  & No\\ \hline
 2 & TTSFW & Yes  & Yes\\ \hline
 3 & PTTW & Yes & No \\ \hline
 4 &\begin{tabular}{@{}c@{}}Full Bandwidth \\ LFM \end{tabular}
	 & No  & Yes \\ \hline
\end{tabular}
\label{tab:table1}
\end{table}

%%%%%%%%%%%%%%%%%%%%%%%%%%%%%%%%%%%%%%%%%%%%%%%%%%%%%%%%%%%%%%%%%%%%%%%%%%%%%%%%%%%%%%
%\newpage
\section{Measurements}
\subsection{Waveform Properties}
\begin{figure} [t!]
	\centering
	\begin{minipage}[t]{0.24\textwidth}
		\centering
		\includegraphics[width=\textwidth]{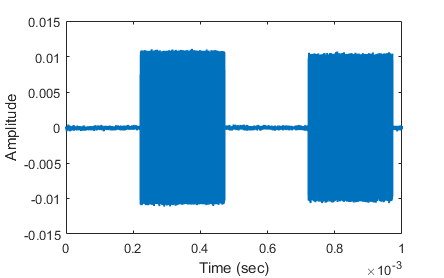}
	\end{minipage}
	\hfill
	\begin{minipage}[t]{0.24\textwidth}
		\centering
		\includegraphics[width=\textwidth]{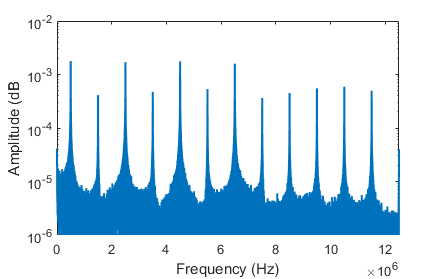}
	\end{minipage}
		\caption{Measured two pulse TTSFW in time domain (left) and in frquency domain (right)}
		\label{fig:test_waveform}
\end{figure}

\begin{figure} [t!]
	\begin{center}
		\includegraphics[width=3.5in]{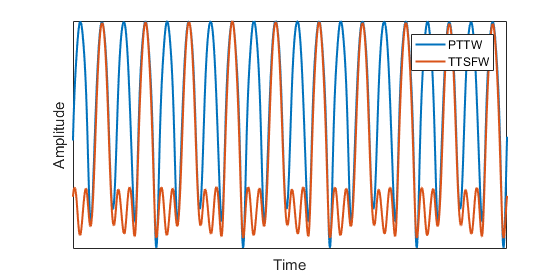}
	\end{center}
	\caption{Measured matched filter output of the PTTW and TTSFW waveforms produced on the X310 using two pulses}
	\label{fig:mf}
\end{figure}

%The properties of the TTSFW have been derived and have been expressed through simulation but still need to verified using equipment. 
A comparison test for the accuracy of the TTSFW and the PTTW was performed using an Ettus X310 software defined radio (SDR) which has an operational bandwidth of 10 MHz -- 6 GHz, and an instantaneous bandwidth of 160 MHz. The SDR interfaces with a host computer though a 10 GHz Ethernet cable and is processed using LabView. Due to processing rate limitations the sampling rate of the SDR is confined to 25 MHz. To ensure that the resulting matched filter was not too discretized both waveforms are taken to have equivalent pulse bandwidth $\Delta f = f_2-f_1$ to be 4 MHz and a pulse duration of 1 ms with duty cycle of 50\%. To ensure that both waveforms have the same integration time of 0.5 ms, the time is split into two equal 0.25ms pulses of the TTSFW. The corresponding bandwidth $\Delta f$ for the TTSFW is 2 MHz which means that every other lobe of the the PTTW is notched out. An image of the TTSFW in the time domain and in the frequency domain can be seen in Fig. \ref{fig:test_waveform} and an image of the comparison of the matched filter outputs can be seen in Fig. \ref{fig:mf}.

Measurements were taken using a loopback architecture, where the transmit channel on the SDR was connected directly to the receive channel via a SMA cable. The received waveform was passed through a matched filter and then interpolated eight times using a built-in LabView spline function to improve the accuracy of the measurement. The interpolation was limited to 8 due to processing limitations and the desire to keep the processing running real-time. Increased latencies resulted in dropped packets, leading to incorrect measurements. 
%because anything higher puts too much strain on the computer processor causing an increasing in the receive loop processing time duration and therefore delaying the next capture the SDR can make. If too much time has elapsed between captures, packet drops between the host computer and SDR occur resulting in system failure. 
The peak of 100 matched filter outputs were recorded and averaged together. The received signal SNR
%The SNR of the received measured signals 
was estimated using an eigenvalue decomposition approach \cite{6860836}. The received signal can be represented by the sampled matrix
\begin{equation}
\textbf{X}=\
\begin{bmatrix}
    \chi_{1,1} & \chi_{1,2} & \dots  & \chi_{1,L} \\
    \chi_{2,1} & \chi_{2,2} & \dots  & \chi_{2,L} \\
    \vdots 		 & \vdots 		& \ddots & \vdots \\
    \chi_{N,1} & \chi_{N,2}  & \dots  & \chi_{N,L} \\
\end{bmatrix}
\end{equation}
where $N$ is the total number of samples per capture and $L$ is then total number of signal observations. From this the covariance matrix can be computed as
\begin{equation}
	\textbf{R}_x=\frac{1}{N}\textbf{X}\textbf{X}^H
\end{equation}
where $\textbf{X}^H$ is the Hermitian of the matrix $\textbf{X}$.

\begin{figure} [t!]
	\begin{center}
		\includegraphics[width=3.5in]{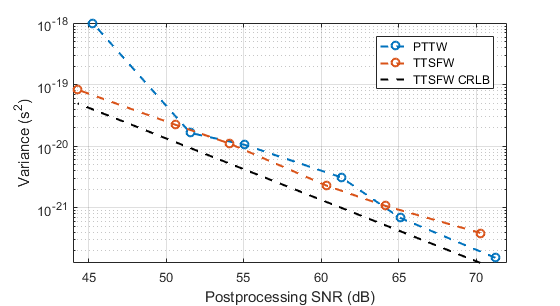}
	\end{center}
	\caption{Comparison of variance of PTTW and TTSFW to theory}
	\label{fig:var}
\end{figure}

Once the covariance matrix is calculated, the eigenvalues $\lambda_l$ are calculated using singular-value decomposition, and the resulting eigenvalues are rank ordered from largest to smallest. Because the various tones in the signal are generated by the same system, they are correlated, yielding a single eigenvalue $\lambda_1$ which will be the largest eigenvalue for non-negative SNR. The remaining $\lambda_2 - \lambda_L$ eigenvalues represent the noise, and therefore the noise level can be estimated by
\begin{equation}
	\gamma^2=\frac{1}{L-1}\sum_{l=2}^L{\lambda_l}
\end{equation}
The signal level then is calculated using 
\begin{equation}
	P_s=\frac{\lambda_1-\gamma^2}{L}
\end{equation}
from which the SNR can be obtained by
\begin{equation}
	\mathrm{SNR}_{dB}=10\log_{10}\left(\frac{P_s}{\gamma^2}\right)
\end{equation}
The processing gain can then be calculated as the time-bandwidth product such that
\begin{equation}
	\mathrm{SNR}_{dB}=10\log_{10}\left(T*BW\right)
\end{equation}
where $T$ is the signal time duration and $BW$ is the noise bandwidth. For the case of $T=0.5$ ms and $BW=12.5$ MHz the resulting processing gain is 38 dB. A comparison of the variance two waveforms vs the derived CRLB for the TTSFW in Eq. \ref{eq:12} can be seen in Fig. \ref{fig:var}.

The variances are comparable indicating that the accuracy obtained by the TTSFW is equivalent to that of the PTTW. The achieved ranging accuracy of $2\times10^{-22}$ s enables coherent transmission up to $9.4$ GHz in a two-way ranging measurement, following from the requirement that the error is within $\frac{\lambda}{15}$ of the coherent transmission signal.

\subsection{Wireless Ranging Measurements}
Wireless ranging measurements were taken on a 24 ft semi-enclosed arch range with a corner reflector or an SDR-based active repeater placed in the middle of the range. Horn antennas were placed on the edge of the arch using a wideband (2-12 GHz) antenna as the transmitter and a narrowband (3.5-5.5 GHz) antenna as the receiver. An Ettus X310 SDR was used to create the TTSFW. The horn antenna have greater directivity at higher frequencies which helped with multipath error from the lab environment. For this reason the center frequency was chosen to be 5.25 GHz due to the limiting bandwidth of the narrowband horns.

\begin{figure} [t!]
	\centering
	\begin{minipage}[t]{0.24\textwidth}
		\centering
		\includegraphics[width=\textwidth]{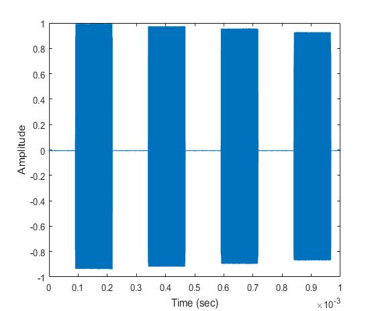}
	\end{minipage}
	\hfill
	\begin{minipage}[t]{0.24\textwidth}
		\centering
		\includegraphics[width=\textwidth]{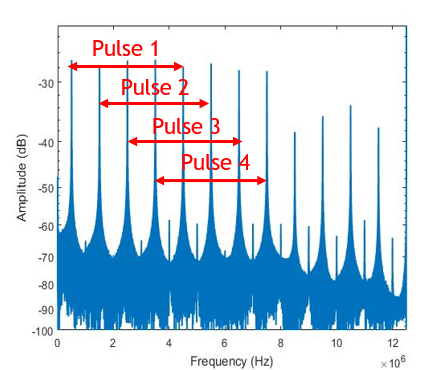}
	\end{minipage}
		\caption{Measured four pulse TTSFW in time domain (left) and in frquency domain (right)}
		\label{fig:waveform}
\end{figure}

\begin{figure} [t!]
	\begin{center}
		\includegraphics[width=3.5in]{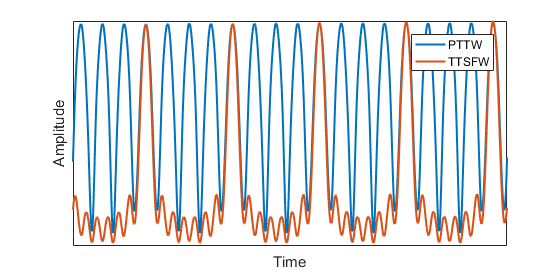}
	\end{center}
	\caption{Measured matched filter output of the PTTW and TTSFW waveforms produced on the X310 using four pulses}
	\label{fig:mf2}
\end{figure}

The waveform that was used was a four-pulse TTSFW with each pulse bandwidth of $\Delta f=4$ MHz and increasing $\delta f=1$ MHz with every consecutive pulse. The sampling frequency was chosen to be 25 MHz which is the highest obtainable stable sample rate with the equipment setup. The total waveform duration was 1 ms, in which each pulse has a duration of 250 $\mu$s and a duty cycle of $50\%$. These waveforms had a preprocessing SNR of approximately 30 dB. An image of the waveform in the time domain and frequency domain can be seen in Fig. \ref{fig:waveform}. The disambiguation properties of the four pulse waveform can be seen in Fig. \ref{fig:mf2}.

The first measurement was taken using a single SDR transmitting to a corner reflector. This experiment is representative of a single slave node where a slave node determines its relative range to a single point in the array. The corner reflector was placed at the far end of the range (15 ft from the transmitter) and moved 10 ft towards the transmitter in 10 inch increments. The return waveform was then passed through the matched filter and interpolated 8 times to improve the accuracy. The peak was then selected by using a simple peak finding operation since the properties of the TTSFW take care of the disambiguation. At each distance 100 matched filter peaks were averaged together.

\begin{figure} [t!]
	\begin{center}
		\includegraphics[width=3.5in]{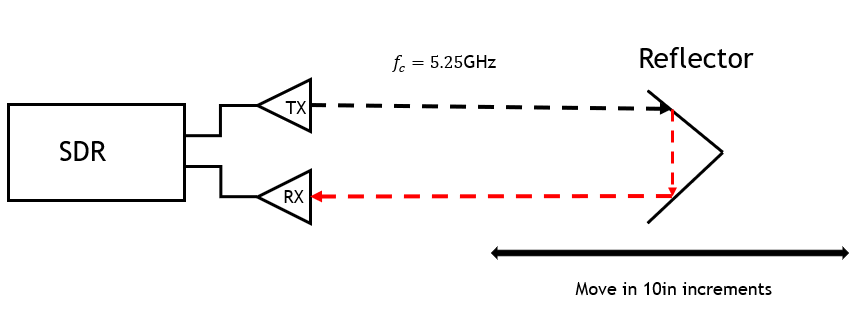}
		
		(a)
		
		\includegraphics[width=3in]{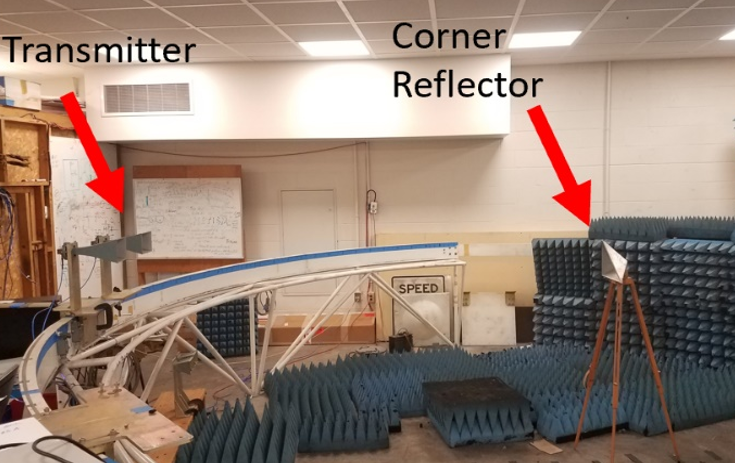}
		
		(b)
		
		\includegraphics[width=3.4in]{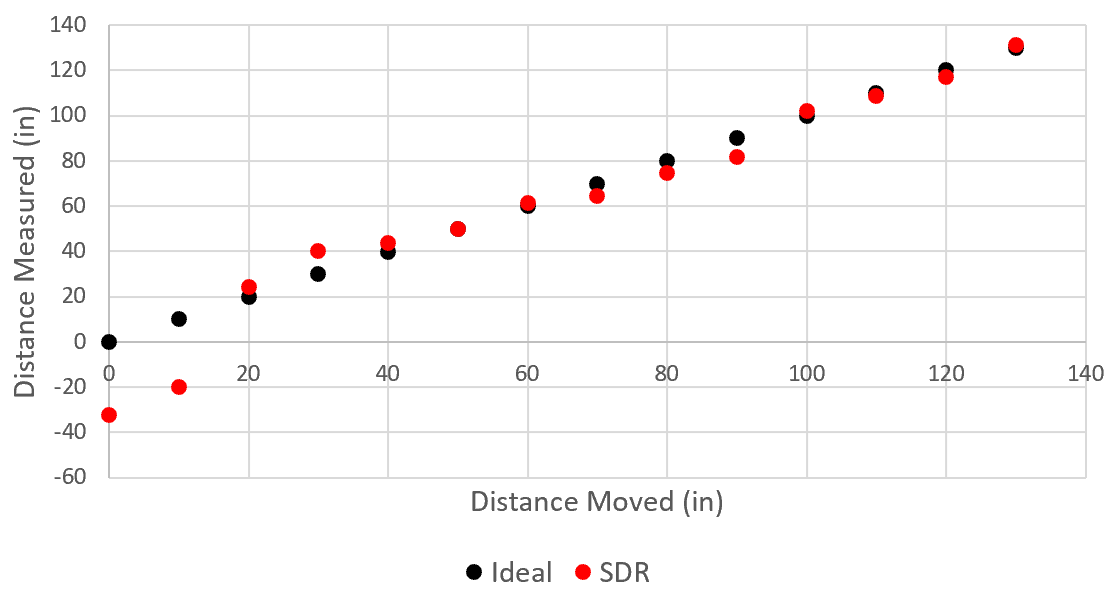}

		(c)
		\includegraphics[width=3.4in]{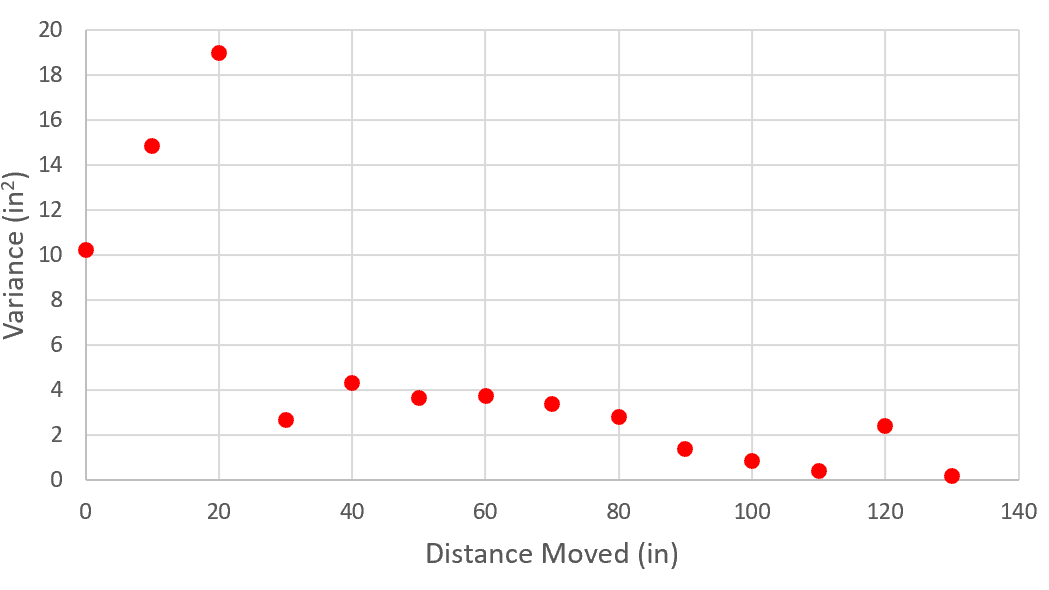}

		(d)
	\end{center}
	\caption{(a) Schematic of single SDR in the arc range (b) Image of experimental setup (c) Range Measurement results (d) Variance of measurement}
	\label{fig:1sdr}
\end{figure}

\begin{figure} [t!]
	\begin{center}
		\includegraphics[width=3.4in]{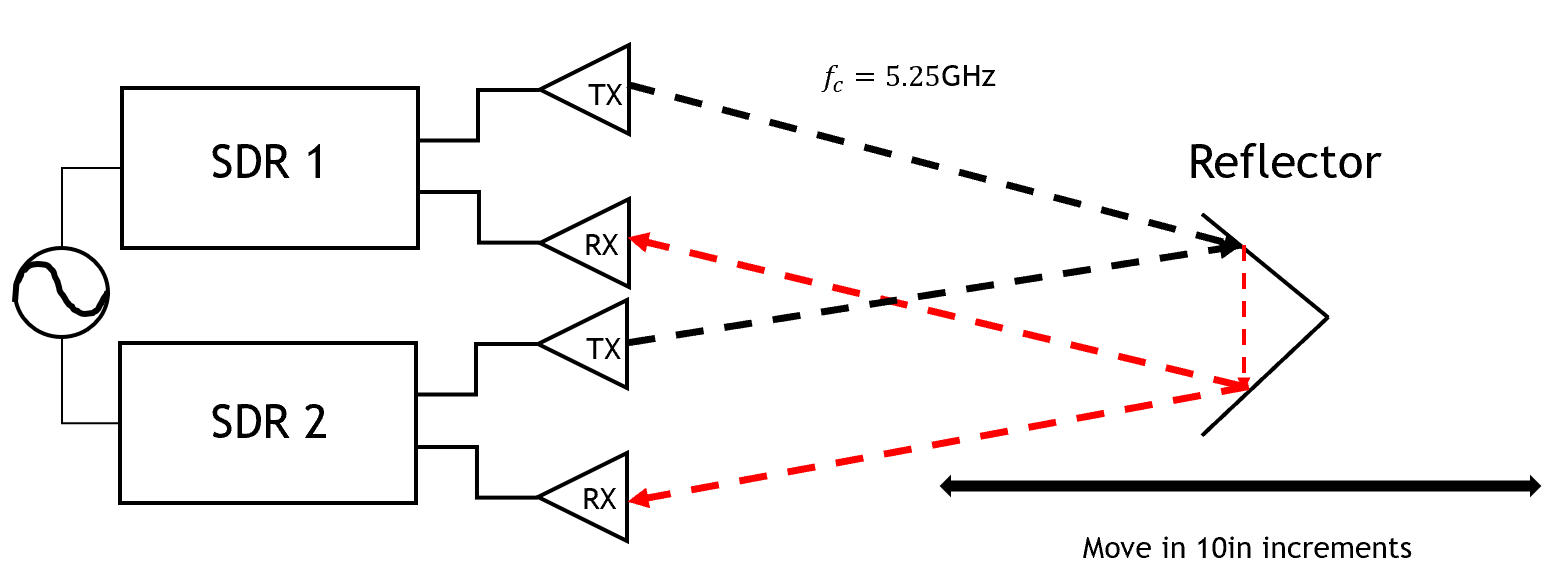}
		
		(a)
		
		\includegraphics[width=3.25in]{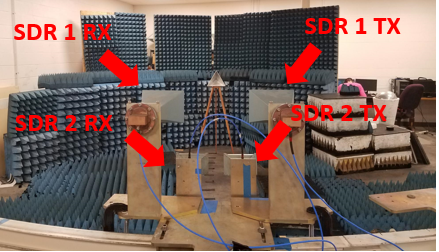}
		
		(b)
		
		\includegraphics[width=3.5in]{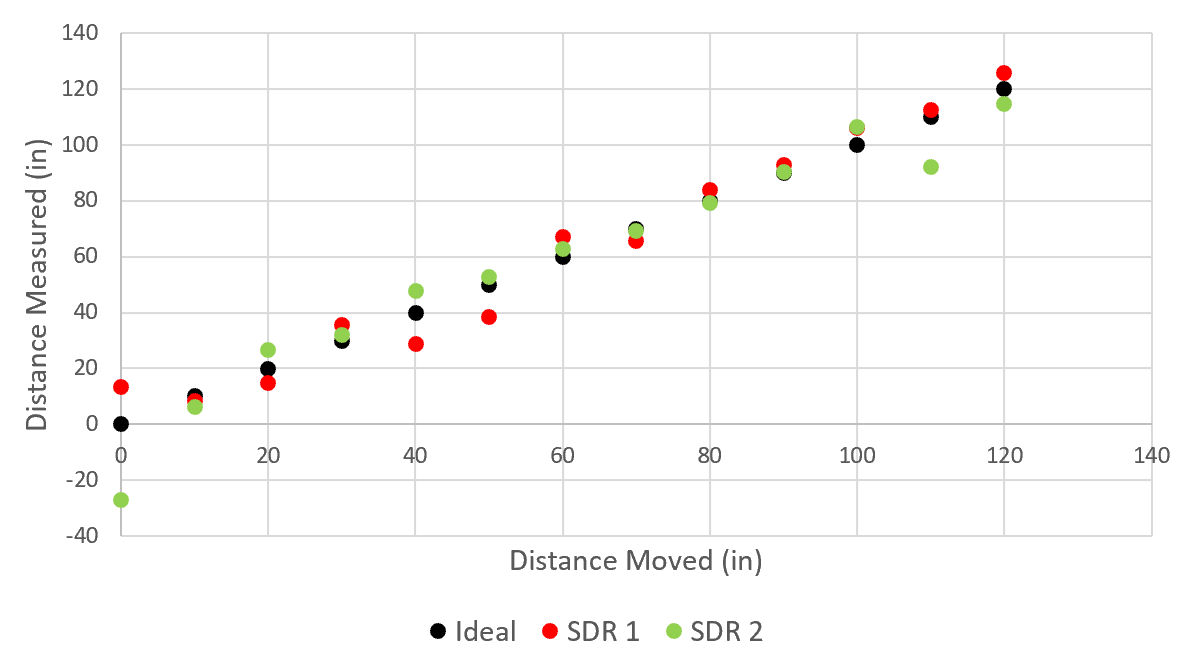}

		(c)
		\includegraphics[width=3.5in]{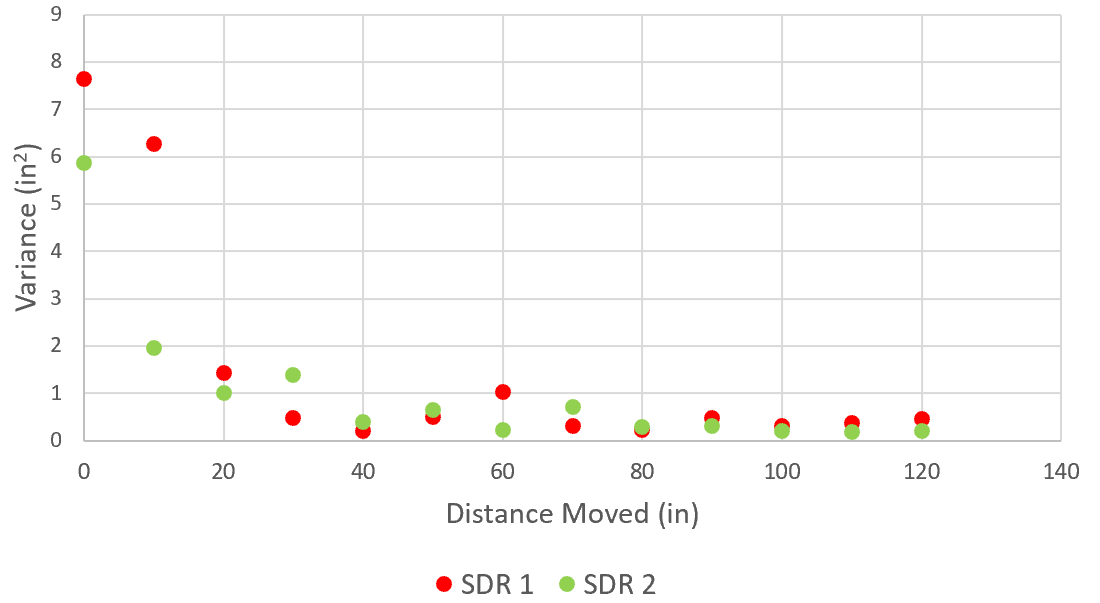}

		(d)
	\end{center}
	\caption{(a) Schematic of two SDRs in the arc range (b) Image of experimental setup (c) Range Measurement results (d) Variance of measurement}
	\label{fig:2sdr}
\end{figure}

After the measurement was taken and the position values were computed, a simple post-processing calibration procedure was implemented to account for any static delays that are inherent to the system. The calibration procedure was performed by taking the average of the differences between the expected value and the measured value at each point and subtracting this average from all points. An image of the performance of a single SDR can be seen in Fig. \ref{fig:1sdr}.

The second measurement was to test the performance of multiple SDRs ranging to a corner reflector simultaneously. This experiment is representative of two slave nodes performing simultaneous ranging. To do this measurement, all of the experiment parameters from the single SDR case remained the same, however two separate SDRs and sets of horn antennas were used. The reference oscillators of the two SDRs were locked via cable; in future work this will be implemented with a wireless link. The two sets of antennas were placed in the same area at different elevations. The waveforms of the two SDRs contained the same set of frequencies but where the first SDR stepeds up in frequency, the second stepped down in frequency, implementing the different step cycles mentioned in the previous section. This resulted in no interference between the waveforms, and in turn allowed for no need for coordination of the waveform start times between the SDRs. An image of the performance of both SDRs can be seen in Fig. \ref{fig:2sdr}.

\begin{figure} [t!]
	\begin{center}
		\includegraphics[width=3.5in]{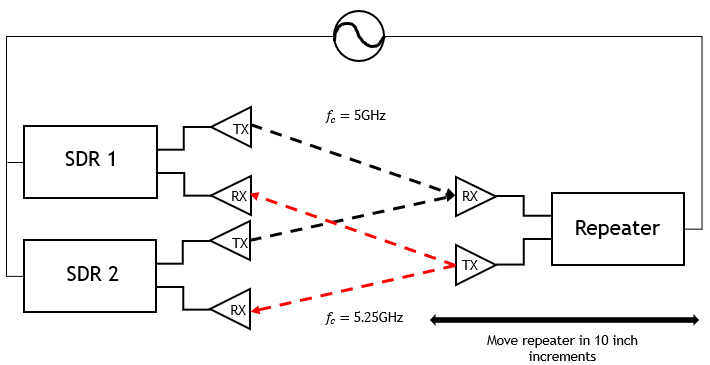}
		
		(a)
		
		\includegraphics[width=3in]{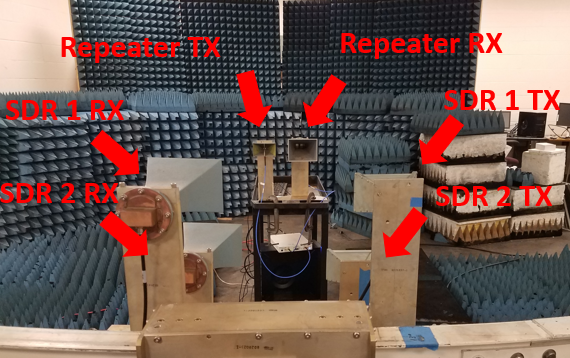}
		
		(b)
		
		\includegraphics[width=3.5in]{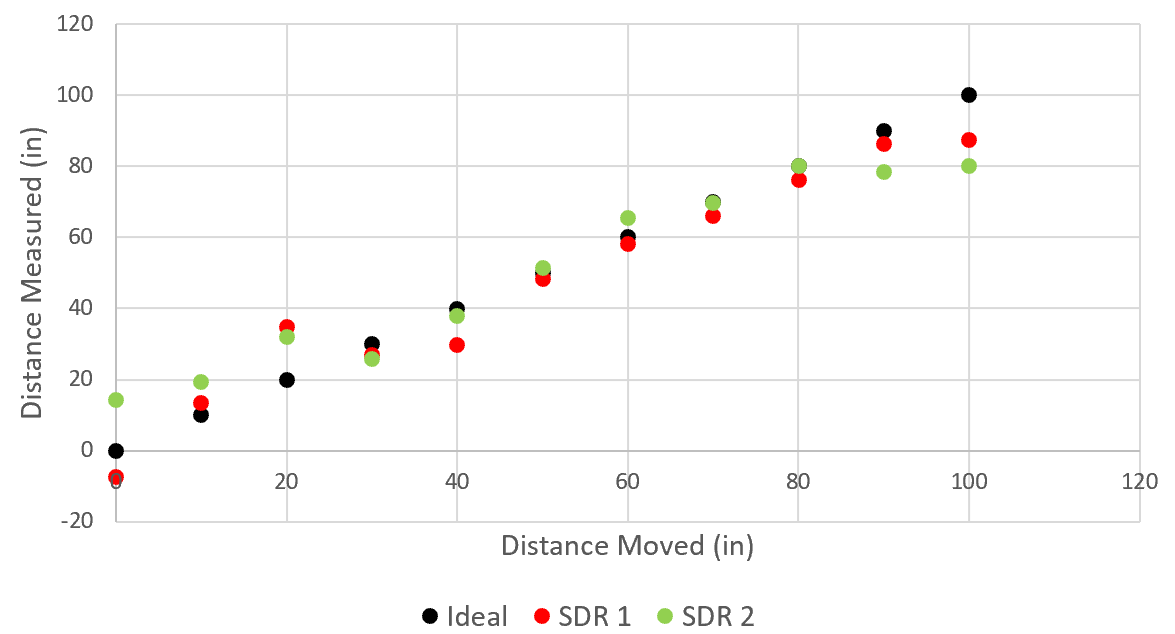}

		(c)
		\includegraphics[width=3.5in]{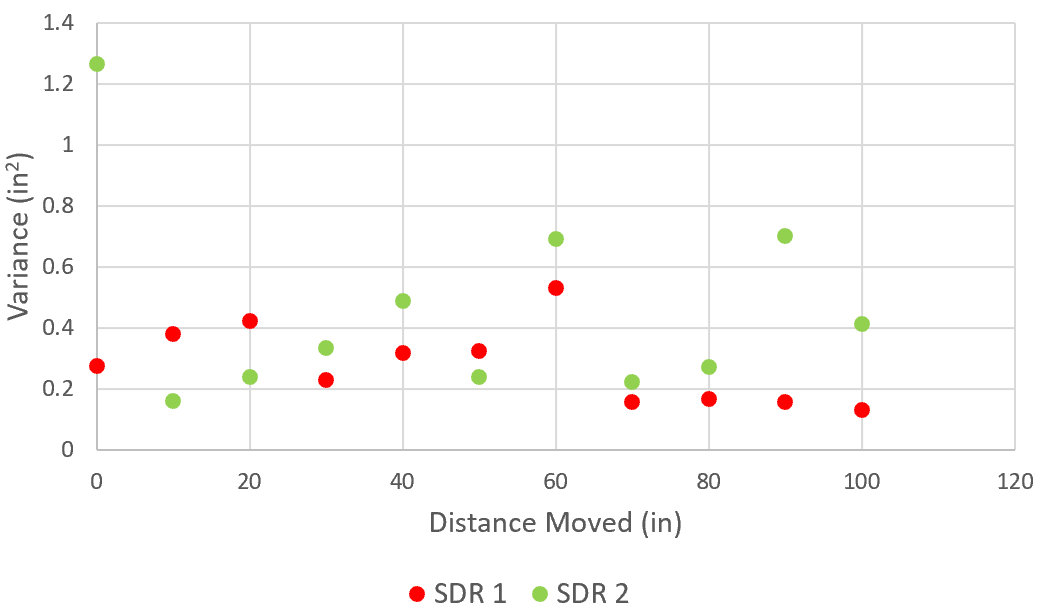}

		(d)
	\end{center}
	\caption{(a) Schematic of three SDRs in the arc range. (b) Image of experimental setup. (c) Range Measurement results. (d) Variance of measurement.}
	\label{fig:3sdr}
\end{figure}

The third measurement was conducted to demonstrate the scalability of the waveform to three separate SDRs. The corner reflector was removed from the setup and replaced by an SDR acting as an active repeater that captured the incoming signals and retransmitted them with increased gain. This helped to increase the SNR of the system by changing the signal decay from $1/r^4$ to $1/r^2$ due to the retransmit gain, and further reduced the effects of multipath. This experiment is representative of two slave nodes ranging to a master node which here is given by the repeater. In a fully distributed array this master node would be sending out the reference coherent action in which the slave nodes, now with knowledge of the distance to the master, will phase adjust their output creating a coherent system. The center frequencies of transmit and receive were separated to also help with multipath issues. The two SDRs on the side of the range were transmitting at 5 GHz and the repeater responded at 5.25 GHz. The performance at each distance can be seen in Fig. \ref{fig:3sdr}. Clearly, the ranging waveform produces accurate range measurements, below 1 in. variance for multiple nodes simultaneously.

%%%%%%%%%%%%%%%%%%%%%%%%%%%%%%%%%%%%%%%%%%%%%%%%%%%%%%%%%%%%%%%%%
\section{Conclusion}
This work has shown that the process of FDM for spectral channeling and a TTSFW can be used as a  methodology for scalable internode ranging for coherent distributed arrays. This waveform was shown to have the measurement accuracy of a PTTW along with the disambiguation properties of the SFW. Although there were errors in the absolute range seen by both SDRs, which are attributed to multipath issues, the relative ranging performance seen by both SDRs is quite good, and does not degrade as more nodes are added to the system, indicating a robust, scalable ranging approach.
%%%%%%%%%%%%%%%%%%%%%%%%%%%%%%%%%%%%%%%%%%%%%%%%%%%%%%%%%%%%%%%%%%%%%%%%%%%%%%%%%%%%%%%%%%%%%%%

\section*{Acknowledgment}
Contractor acknowledges Government's support in the publication of this paper. This material is based upon work funded by AFRL under AFRL Contract No. FA8750-17-C-0182 and by The Defense Advanced Research Projects Agency under grant number N66001-17-1-4045.

\bibliographystyle{IEEEtran}
\bibliography{IEEEabrv,Scalable_Ranging_-_Revision_2_-_Final}

\end{document}